\documentclass[aps, twocolumn, pra, showpacs, superscriptaddress]{revtex4}
\usepackage{amsmath}
\usepackage{amsfonts}
\usepackage{mathrsfs}
\usepackage{graphicx}
\usepackage{dcolumn}
\usepackage{bm}

\begin{document}

\title{Quantum criticality of a one-dimensional Bose-Fermi mixture}
\author{Xiangguo Yin}
\affiliation{Beijing National Laboratory for condensed matter physics, Institute of
Physics, Chinese Academy of Sciences, Beijing 100190, China}
\author{Xi-Wen Guan}
\email{xwe105@rsphysse.anu.edu.au}
\affiliation{Department of Theoretical Physics, Research School of Physics and
Engineering, Australian National University, Canberra ACT 0200, Australia}
\author{Yunbo Zhang}
\affiliation{Department of Physics and Institute of Theoretical Physics, Shanxi
university, Taiyuan 030006, China}
\author{Shu Chen}
\email{schen@aphy.iphy.ac.cn}
\affiliation{Beijing National Laboratory for condensed matter physics, Institute of
Physics, Chinese Academy of Sciences, Beijing 100190, China}

\begin{abstract}
The one-dimensional interacting Bose-Fermi mixtures, exhibiting quantum
phase transitions at zero temperature, are particularly valuable for the
study of quantum critical phenomena. In the present paper, we analytically
study quantum phase diagram, equation of state and quantum criticality of
the Bose-Fermi mixture using the thermodynamic Bethe ansatz equations. We
show that thermodynamical properties display universal scaling behaviour at
quantum criticality. Furthermore, quantum criticality of the Bose-Fermi
mixture in an harmonic trap is also studied within the local density
approximation. We thus demonstrate that the phase diagram and critical
properties of the bulk system provide insights into understanding universal
features of many-body critical phenomena.
\end{abstract}

\date{\today}
\maketitle


\section{Introdunction}

Mixtures of ultracold bosonic and fermionic atoms have attracted intensive
studies both experimentally and theoretically \cite{Hulet,BF_exp,Fukuhara}.
By loading cold atoms in one-dimensional (1D) waveguides and tuning the
effective interactions by Feshbach resonance, it is possible to simulate
striking quantum many-body phenomena in 1D strongly correlated systems in
the whole regime of interaction strength \cite{Olshanii,Gorlitz,Moritz}. The
exquisite tunability with ultracold atoms confined to low dimensions has
provided unprecedented opportunities for investigating and testing the
theory of exactly solvable many-body systems \cite%
{Paredes,Kinoshita,Amerongen,Haller,Liao}. These include remarkable
experimental progresses in the realization of Tonks-Girardeau gas \cite%
{Paredes,Kinoshita}, super-Tonks-Girardeau gas \cite{Haller}, Yang-Yang
thermodynamics for ultra-cold Bose gas of $^{87}$Rb \cite{Amerongen}, and
the exotic density profiles of the attractive Fermi gas in an harmonic trap
\cite{Liao}. The current experimental progresses are capable of simulating
1D Bose-Fermi mixtures.

Recently, various theoretical methods have been used to study quantum phases
and correlations of the 1D Bose-Fermi mixtures, such as the mean-field
approach \cite{Das} and Tomonaga-Luttinger liquid (TLL) theory \cite%
{Cazalilla,Mathey}, etc. The 1D Bose-Fermi mixture with equal masses of
bosons and fermions and with the same strength of delta-function interaction
between boson-boson and boson-fermion is exactly solvable \cite{Lai}. This
model has been attracted a renewed interest \cite%
{Imambekov,Batchelor,Frahm,Yin} due to the experimental progress with cold
atomic systems.

Particular theoretical interest has been paid to the ground-state properties
at zero temperature \cite{Imambekov,Batchelor,Frahm}. However, there are
very few studies on the thermodynamics and quantum critical phenomena of the
model. In review of the realistic cold atomic systems trapped in external
potentials at finite temperatures, it is significantly important to
understand how to unambiguously determine the zero temperature phase diagram
from the knowledge of finite temperature quantities of trapped gases. In the
1D mixture of quantum gases, true quantum phase transitions occur as the
driving parameters vary across the phase boundaries at zero temperature,
such as chemical potential, magnetic field and densities, etc. In
particular, 1D quantum critical phenomena associated with quantum phase
transitions at zero temperature give physical origin of quantum criticality
\cite{Fisher,Sachdev} and provide an insight into understanding of universal
scaling behaviour of thermodynamical properties in quantum critical regimes
\cite{Zhou,Cai,Guan,GuanJPA}. By using the universal scaling functions, it
has been demonstrated that the zero-temperature phase diagrams of various
systems can be mapped out from the finite-temperature density profiles \cite%
{Zhou,Wang,Guan}. Most recently, the high-resolution imaging techniques have
allowed to measure the density profiles and density fluctuations of the
trapped atomic gases very precisely \cite{Ho,Chin,Greiner,Sherson,Huang} and
thus provide essential tools to study quantum phase transitions and quantum
criticality.

In general, quantum fluctuations are strongly coupled with thermal
fluctuations in the quantum critical regime. Therefore, quantum criticality
is among the most challenging problems in condensed matter physics. In order
to extract correct universal scaling functions which control proper thermal
and quantum fluctuations at quantum criticality, high precision of finite
temperature thermodynamics is desirable. Usually, accessing to the
thermodynamic properties of integrable models at finite temperatures is
notoriously difficult and presents a formidable challenge in theoretical and
mathematical physics. In the present paper, we analytically determine the
zero temperature phase diagram of the integrable Bose-Fermi mixture. We
further derive equation of state and explore universal scaling behaviour of
thermodynamical properties at quantum criticality using the thermodynamical
Bethe ansatz (TBA) equations. Using exact analytical result obtained, we
also demonstrate that the zero temperature phase diagram and quantum
criticality can be mapped out from finite temperature density profiles of
the trapped gas within the local density approximation.

The paper is organized as follows. In section \ref{Section2}, we present the
TBA equations for the model and analytically determine the phase diagram of
the Bose-Fermi mixture at zero temperature. In section \ref{Section3}, we
derive equation of state and explore the universal scaling behavior of the
density and compressibility near the critical points. In section \ref%
{Section4}, the quantum criticality of the gas in an harmonic trap is
studied within the local density approximation. A conclusion is presented in
the section \ref{Section5}.

\section{Model and phase diagram at zero temperature}

\label{Section2}

We consider a 1D interacting Bose-Fermi mixture described by the Hamiltonian
\begin{align}
\hat{H}& =\int_{0}^{L}dx\left( \frac{\hbar ^{2}}{2m_{b}}\partial _{x}\Psi
_{b}^{\dag }\partial _{x}\Psi _{b}+\frac{\hbar ^{2}}{2m_{f}}\partial
_{x}\Psi _{f}^{\dag }\partial _{x}\Psi _{f} + \right.  \notag \\
& \left. \frac{g_{bb}}{2} \Psi _{b}^{\dag }\Psi _{b}^{\dag }\Psi _{b}\Psi
_{b}+g_{bf}\Psi _{b}^{\dag }\Psi _{f}^{\dag }\Psi _{f}\Psi _{b} - \mu_f \Psi
_{f}^{\dag }\Psi _{f} - \mu_b \Psi _{b}^{\dag }\Psi _{b}\right) ,
\label{H_second form}
\end{align}%
where $\Psi _{b}$, $\Psi _{f}$ are boson and fermion field operators, $m_{b}$%
, $m_{f}$ are the masses, $\mu_{b}$, $\mu_{f}$ are chemical potentials of
bosons and fermions, and $g_{bb}$, $g_{bf}$ are boson-boson and
boson-fermion interaction strengths, respectively. Here we consider the
fully-polarized fermions, therefore, the Pauli principle excludes their
s-wave interaction ($g_{ff}=0$). This model is exactly solvable \cite%
{Lai,Imambekov} for equal masses and equal boson-boson and boson-fermion
interaction strengths, i.e., $m_{b}=m_{f}=m$ and $g_{bb}=g_{bf}=g$. For our
convenience, we assume $\hbar =2m=1$. The first quantization form of the
exactly solvable Hamiltonian (\ref{H_second form}) can be written as
\begin{equation}
\hat{H}=-\sum_{i=1}^{N}\frac{\partial ^{2}}{\partial x_{i}^{2}}+2c\sum_{i<j}
\delta \left( x_{i}-x_{j}\right) - \mu N - \frac{H}{2}\left(N_f -N_b \right)
\label{H_first quantization}
\end{equation}%
with $c=mg/\hbar ^{2}$. Here the particle number $N=N_{b}+N_{f}$ with $N_{b}$
bosons and $N_{f}$ fermions. The chemical potential $\mu $ and the effective
magnetic field $H$ are defined as $\mu = \left( \mu _{f}+\mu _{b}\right) /2$
and $H=$ $\mu _{f}-\mu _{b}$. 
The many-body wave function is supposed to be symmetric with respect to
indices $i=\left\{ 1,2,...,N_{b}\right\} $ (bosons) and antisymmetric with
respect to $\left\{ N_{b}+1,N_{b}+2,...,N_{f}\right\} $ (fermions). Thus the
$N$-body wave function can be written as $N!\times N!$ superpositions of
individual plane waves associated with $N$ quasi-momenta $k_i$ with $%
i=1,\ldots, N$ by means of the Bethe ansatz \cite{Lai,Imambekov}.

The spectrum of the system is given by $E=\sum_{i=1}^{N}k_{i}^{2}$, where
the quasi-momenta $k_{i}$ is subject to the so-called Bethe ansatz equations
(BAEs) \cite{Lai,Imambekov}. In thermodynamic limits, i.e. $N\rightarrow
\infty ,L\rightarrow \infty $ and $N/L$ is finite, and at finite
temperatures, the equilibrium states become degenerate. Yang and Yang \cite%
{Yang} showed that true physical states in integrable systems can be
determined from the minimization conditions of Gibbs free energy subject to
the Bethe ansatz equations. Accordingly, in the thermodynamics limit with $%
N_{b}/L$ and $N_{f}/L$ fixed, minimization of the Gibbs free energy gives
rise to the following nonlinear integral equations \cite{Yin}, i.e., the TBA
equations for the integrable Bose-Fermi mixture: ($k_{B}=1$)
\begin{align}
& \epsilon \left( k\right) =k^{2}-\mu _{f}  \notag \\
& -T\int_{-\infty }^{\infty }a_{1}\left( \Lambda -k\right) \ln \left( 1+\exp
\left( -\varphi \left( \Lambda \right) /T\right) \right) d\Lambda ,  \notag
\\
& \varphi \left( \Lambda \right) =\mu _{f}-\mu _{b}  \notag \\
& -T\int_{-\infty }^{\infty }a_{1}\left( k-\Lambda \right) \ln \left( 1+\exp
\left( -\epsilon \left( k\right) /T\right) \right) dk,  \label{TBA_nonlinear}
\end{align}%
where $a_{\ell }\left( x\right) =\frac{1}{2\pi }\frac{\ell c}{(\ell
c)^{2}/4+x^{2}}$. For fixed temperature $T$ and chemical potential $\mu _{f}$%
, $\mu _{b}$, the pressure is given by
\begin{equation}
p=\frac{T}{2\pi }\int_{-\infty }^{\infty }\ln \left( 1+\exp \left( -\frac{%
\epsilon \left( k\right) }{T}\right) \right) dk.  \label{energy}
\end{equation}%
The particle density and compressibility for fermions and bosons, and the
entropy per length can be obtained from
\begin{eqnarray}
n_{i} &\equiv &N_{i}/L=\partial p/\partial \mu _{i},\text{ \ }i=f,b
\label{particle_density} \\
\kappa _{i} &\equiv &\partial n_{i}/\partial \mu ,\text{ \ \ \ \ \ \ \ \ \ }%
i=f,b \\
S/L &\equiv &\partial p/\partial T.\text{ \ \ }
\end{eqnarray}%
%
%
%
%
\begin{figure}[tbp]
\includegraphics[width=1.0\linewidth]{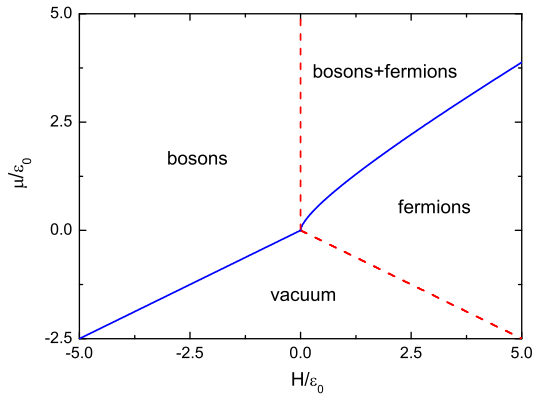}
\caption{(Color online) Phase diagram in $\protect\mu -H$ plane. Three
distinguished phases are resulted from varying the chemical potential and
magnetic field, i.e. pure boson phase for $H<0$ and $\protect\mu >H/2$; pure
fermion phase below the phase boundary (\protect\ref{boundary}) in the
region $\protect\mu >-H/2$; and the mixture of bosons and fermions above the
phase boundary (\protect\ref{boundary}) in the region $H>0$. }
\label{f1}
\end{figure}

The phase diagram of the Bose-Fermi mixture can be analytically determined
from the TBA equations (\ref{TBA_nonlinear}) in the zero temperature limit.
As shown in Fig. \ref{f1}, the phase diagram consists of three quantum
phases: pure bosons, pure fermions and the mixture of bosons and fermions
except the vacuum, which are separated by four boundary lines with condition
$n_{f}=0$ or $n_{b}=0$. In order to obtain the full phase diagram of the
Bose-Fermi mixture, we introduce two sets of the TBA equations with
different reference states \cite{Li}. The TBA equations based on the
fermionic reference state determine the two boundary lines for $H>0$,
whereas the TBA equations based on the bosonic reference state determine the
two boundary lines for $H\le0$.

In the limit $T\rightarrow 0$, the TBA equations (\ref{TBA_nonlinear}) based
on the fermionic reference state reduce to
\begin{align}
& \epsilon \left( k\right) =k^{2}-\mu _{f}+\int_{-\Lambda _{F}}^{\Lambda
_{F}}a_{1}\left( \Lambda -k\right) \varphi ^{-}\left( \Lambda \right)
d\Lambda ,  \notag \\
& \varphi \left( \Lambda \right) =\mu _{f}-\mu
_{b}+\int_{-k_{F}}^{k_{F}}a_{1}\left( k-\Lambda \right) \epsilon ^{-}\left(
k\right) dk,  \label{TBA_zero}
\end{align}%
where the dressed energies $\epsilon ^{-}(k)$ and $\varphi ^{-}(\Lambda )$
correspond to the occupied states for $k\in \lbrack -k_{F},k_{F}]$ and $%
\Lambda \in \lbrack -\Lambda _{F},\Lambda _{F}]$, respectively. The positive
parts of the dressed energies $\epsilon (k)$ and $\varphi (\Lambda )$
corresponds to the unoccupied states. The integration boundaries $k_{F}$ and
$\Lambda _{F}$ characterize the Fermi points with the conditions $\epsilon
^{-}\left( \pm k_{F}\right) =\varphi ^{-}\left( \pm \Lambda _{F}\right) =0$.
The pressure is given by
\begin{equation}
p=-\frac{1}{2\pi }\int_{-k_{F}}^{k_{F}}\epsilon ^{-}\left( k\right) dk
\end{equation}%
at zero temperature. We calculate the particle densities through the
relations (\ref{particle_density}). Nevertheless, the boundary lines, which
correspond to $n_{f}=0$ or $n_{b}=0$, can be determined by analyzing the
dressed energy at the point $k_{F}=0$ or $\Lambda _{F}=0$.

For a pure fermion state, $\varphi \left( \Lambda \right) $ is
gapfull, i.e.
$\varphi \left( \Lambda \right) >0 $. Thus the TBA equations (\ref{TBA_zero}%
) reduce to the free fermion potential $\epsilon \left( k\right) =k^{2}-\mu
_{f}$. The phase boundaries for the phase transitions from vacuum into the
pure fermion state and from pure fermion state into the mixture of bosons
and fermions are determined by
\begin{eqnarray}
\epsilon ^{-}\left( 0\right) &\leq &0,  \label{fermion-1} \\
\varphi ^{-}\left( 0\right) &\leq &0  \label{boson-2}
\end{eqnarray}%
respectively. From the condition (\ref{fermion-1}), we obtain the phase
boundary for pure fermions $\mu _{f}\geq 0$, or equivalently $\mu \geq -H/2$.

From the conditions (\ref{boson-2}) and $\epsilon \left( k\right) =k^{2}-\mu
_{f}$, we obtain the phase boundary
\begin{equation}
\tilde{H}\leq \frac{1}{2\pi }\left[ \left( 4\tilde{\mu}_{f}\allowbreak
+1\right) \arctan \sqrt{4\tilde{\mu}_{f}}-\allowbreak \sqrt{4\tilde{\mu}_{f}}%
\right]  \label{boundary}
\end{equation}%
for the coexistence of bosons and fermions. Here we used the dimensionless
units $\tilde{H}\equiv H/\varepsilon _{0}$ and $\tilde{\mu}_{f}\equiv \mu
_{f}/\varepsilon _{0}$ with $\varepsilon _{0}=c^{2}$. For strong coupling
regime, i.e. $\mu _{f}\ll \varepsilon _{0}$, or $H\ll \varepsilon _{0}$, we
find phase boundary condition (\ref{boundary}) reduces to
\begin{equation}
\tilde{\mu}\geq \frac{1}{4}\left( \left( 3\pi \tilde{H}\right) ^{2/3}\left(
1+\frac{2}{15}\left( 3\pi \tilde{H}\right) ^{2/3}\right) -2\tilde{H}\right)
\end{equation}%
with $\tilde{\mu}\equiv \mu /\varepsilon _{0}$. For weak coupling regime,
i.e. $\mu _{f}\ll \varepsilon _{0}$, or $H\gg \varepsilon _{0}$, the phase
boundary condition becomes
\begin{equation}
\tilde{\mu}\geq \frac{\tilde{H}}{2}+\frac{2}{\pi }\sqrt{\tilde{H}}-\frac{1}{4%
}\left( 1-\frac{8}{\pi ^{2}}\right) .
\end{equation}

On the other hand, at zero temperature, the TBA equations with the Bose
state as the reference state is given by \cite{Li}
\begin{eqnarray}
\epsilon \left( k\right) &=&k^{2}-\mu _{b}+\int_{-k_{F}}^{k_{F}}a_{2}\left(
k-k^{\prime }\right) \epsilon ^{-}\left( k^{\prime }\right) dk^{\prime }
\notag \\
&&+\left( \int_{-\infty }^{-\Lambda _{F}}+\int_{\Lambda _{F}}^{\infty
}\right) a_{1}\left( k-\Lambda \right) \varphi ^{-}\left( \Lambda \right)
d\Lambda  \notag \\
\varphi \left( \Lambda \right) &=&\mu _{b}-\mu
_{f}-\int_{-k_{F}}^{k_{F}}a_{1}\left( k-\Lambda \right) \epsilon ^{-}\left(
k\right) dk  \label{TBA_zero2}
\end{eqnarray}%
with the Fermi points $\epsilon ^{-}\left( \pm k_{F}\right) =\varphi
^{-}\left( \pm \Lambda _{F}\right) =0$. Similarly, the dressed energies $%
\epsilon^-(k)$ and $\varphi^-(\Lambda)$ correspond to the occupied states
for $k\in \lbrack -k_{F},k_{F}]$ and $\Lambda \in \lbrack -\Lambda
_{F},\Lambda _{F}]$, respectively. We see that for the Bose reference state
the fully-polarized fermions provide a ferromagnetic ordering at the ground
state. If $H<0$, i.e. $\mu_f< \mu_b$, the dressed energy $\varphi (\Lambda) $
is greater than zero. The dressed energy $\varphi$ is gapful. Thus the
ground state is a pure boson state. Therefore, for $H<0$, the TBA equations (%
\ref{TBA_zero2}) reduce to Yang-Yang thermodynamics equations for the
Lieb-Liniger Bose gas \cite{Yang}
\begin{equation*}
\epsilon \left( k\right) =k^{2}-\mu _{b}+\int_{-k_{F}}^{k_{F}}a_{2}\left(
k-k^{\prime }\right) \epsilon ^{-}\left( k^{\prime }\right) dk^{\prime }
\end{equation*}%
from which we easily determine the phase boundary for the phase transition
from vacuum into the pure boson state, i.e., $\mu \geq H/2$, see Figure \ref%
{f1}.

\section{Equation of state and universal scalings}

\label{Section3} \bigskip

Recent experiments on quantum criticality of ultracold atoms \cite%
{Chin,Huang} and theoretical scheme of mapping out quantum
criticality of cold atoms \cite{Zhou,Cai,Guan,GuanJPA} open the
possibility to explore such universal behavior in low dimensional
multi-component interacting Fermi and Bose gases. As the
temperature is tuned over the crossover temperatures the scaling
functions of thermodynamical properties give rise to universal
behavior which entirely depends on the symmetry of excitation
spectrum and dimensionality of the system. This gives a promising
way to explore the hidden symmetry of these models, for example,
the quantum Ising model with transverse file displays emergent E8
symmetry \cite{Exp8}. In the critical regime, the thermodynamic
functions of the homogeneous gas can be cast into some universal
scaling forms \cite{Fisher,Sachdev}. For example, the density and
compressibility near the critical point $\mu =\mu _{c}$ can be
written as
\begin{equation}
n\left( T,\mu \right) =n_{0}+T^{\frac{d}{z}+1-\frac{1}{\nu z}}\mathcal{G}%
\left( \frac{\mu -\mu _{c}}{T^{\frac{1}{\nu z}}}\right) ,  \label{us1}
\end{equation}%
\begin{equation}
\kappa \left( T,\mu \right) =\kappa _{0}+T^{\frac{d}{z}+1-\frac{2}{\nu z}}%
\mathcal{F}\left( \frac{\mu -\mu _{c}}{T^{\frac{1}{\nu z}}}\right) .
\label{us2}
\end{equation}%
Here $n_{0}$ ($\kappa _{0}$ )is the regular part of the density
(compressibility) induced from the background, $\mathcal{G}$ ($\mathcal{F}$)
is a universal scaling function describing the singular part of the density
(compressibility) near the critical point $\mu _{c}$, $d$ is the
dimensionality of the system, $z$ is the dynamical critical exponent and $%
\nu $ is the correlation length exponent. From the above relation, the
dimensionless universal scaling functions $\mathcal{G}(\frac{\mu -\mu _{c}}{%
T^{\nu z}})$ and $\mathcal{F}(\frac{\mu -\mu _{c}}{T^{\nu z}})$ display
universal scaling behaviour near the critical point of $\mu =\mu _{c}$, i.e.
the density(compressibility) curves with a subtraction of the background
density (compressibility) intersect at the critical point for different
temperatures. This feature can be used to detect the phase boundaries at
zero temperature from finite temperature density profiles of the trapped gas.

Before discussing the universal scaling behavior of the Bose-Fermi mixture,
we will discuss quantum criticality of several simple examples. The simplest
example is the 1D ideal Fermi gas. The ideal Fermi gas obeys the Fermi-Dirac
distribution, it is easy to derive the density distribution for free
fermions $n_{f}(T,\mu )=-\sqrt{T}/\left( 2\sqrt{\pi }\right) Li_{1/2}\left(
-\exp \left( \frac{\mu _{f}}{T}\right) \right) $, where $Li_{n}\left(
x\right) =\sum_{l=1}^{\infty }x^{l}/l^{n}$ is the standard polylogarithm
function. In comparison with the universal scaling Eq.(\ref{us1}), the
critical exponent $z=2$ and the correlation length exponent $\nu =1/2$ with
the dimensionality $d=1$ can be read off the universal scaling form. Here
the scaling function is $\mathcal{G}\left( x\right) =-1/\left( 2\sqrt{\pi }%
\right) Li_{1/2}\left( -\exp \left( x\right) \right) $. There is no
background density for the vacuum-Fermi-gas transition, i.e. $n_{0}=0$,
where the critical point $\mu _{c}=0$. It was also shown that $d=1$, $z=2$
and $\nu =1/2$ for the vacuum into the TLL phase transition in the 1D
hard-core bosons \cite{Zhou}, the 1D attractive Fermi gas \cite{Guan} and
the 1D interacting Bose gas with strongly repulsive interactions \cite%
{GuanJPA}.

\begin{figure}[tbp]
\includegraphics[width=1.10\linewidth]{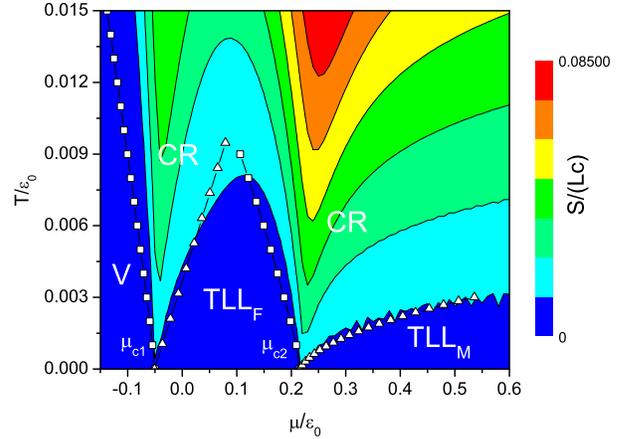}
\caption{(Color online) Contour plot of entropy $S$ vs chemical potential
from the TBA (\protect\ref{TBA_nonlinear}): quantum criticality driven by
chemical potential for $H=0.1\protect\varepsilon _{0}$. The crossover
temperatures (white squares and triangles) separating Vacuum, $TLL_{F}$ and $%
TLL_{M}$ from the quantum critical regimes are determined from the breakdown
of linear temperature-dependent entropy from (\protect\ref{LL}). Here the $%
TLL_{F}$ stands for the TLL of fermions where exponentially small number of
bosons are populated at finite temperatures. Whereas $TTL_{M}$ denotes the
TLL of the mixture. The vacuum evolves into a quasi-classical regime at
finite temperature. }
\label{fig:S}
\end{figure}

In order to investigate the quantum critical behavior in the vicinity of
phase boundary between the phase of a mixture of bosons and fermions and the
phase of the fully-polarized fermions, we will derive the equation of state
from the TBA equations (\ref{TBA_nonlinear}). For strong interacting regime,
i.e. $c\gg 1$, or $T/\varepsilon _{0}\gg 1$, we can rewrite the TBA
equations (\ref{TBA_nonlinear})
\begin{eqnarray}
\epsilon \left( k\right) &\approx &k^{2}/\beta -A  \label{TBA_large} \\
\varphi \left( \Lambda \right) &=&\mu _{f}-\mu _{b}-\frac{4pc}{%
c^{2}+4\Lambda ^{2}}+\frac{4c\left( 4c^{2}-48\Lambda ^{2}\right) p_{2}}{%
\left( c^{2}+4\Lambda ^{2}\right) ^{3}},  \notag
\end{eqnarray}%
where
\begin{eqnarray}
A &\approx &\mu _{f}+T\int_{-\infty }^{\infty }a_{1}\left( \Lambda \right)
\ln \left( 1+\exp \left( -\frac{\varphi \left( \Lambda \right) }{T}\right)
\right) d\Lambda ,  \notag \\
\beta &=&1-\frac{2Tc}{\pi }\int_{-\infty }^{\infty }\frac{4c^{2}-48\Lambda
^{2}}{\left( c^{2}+4\Lambda ^{2}\right) ^{3}}\ln \left( 1+e^{-\frac{\varphi
(\Lambda )}{T}}\right) d\Lambda ,  \notag \\
p_{2} &=&-\frac{\beta T^{2}}{4\sqrt{\pi }}\mathrm{Li}_{\frac{5}{2}}\left(
-e^{\frac{A}{T}}\right) .  \label{p2}
\end{eqnarray}%
With the help of these relations, the pressure (\ref{energy}) can be
calculated in a straightforward way
\begin{equation}
p=-\sqrt{\frac{\beta }{4\pi }}T^{\frac{3}{2}}\mathrm{Li}_{\frac{3}{2}}\left(
-e^{\frac{A}{T}}\right)  \label{P}
\end{equation}%
that serves as the equation of state of the model with strong repulsion. The
thermodynamical properties can be obtained from the usual thermodynamical
relations. This analytical equation of state (\ref{P}) essentially cover the
universal TLL thermodynamics and encode the critical exponents in the
critical regimes.

At very low temperatures, i.e. $T\ll \varepsilon _{0}$, the thermodynamics
of the model is governed by the TLL physics associated with a linear
dispersion. In the mixed phase of bosons and fermions, the low energy
physics belongs to a universality class of a two-component TLL \cite%
{Frahm,Batchelor}. In this low temperature limit, we further calculate the
pressure
\begin{equation*}
p\approx \frac{2A^{3/2}}{3\pi }\left( 1+\frac{\pi ^{2}}{8}\frac{T^{2}}{A^{2}}%
\right) ,
\end{equation*}%
where
\begin{eqnarray}
A &\approx &\mu +\frac{H}{2}-\frac{4}{\pi }\left( \frac{H}{2}-\frac{p}{c}%
\right) \tan ^{-1}\frac{2\Lambda _{0}}{c}  \notag \\
&&+\frac{2H\Lambda _{0}}{\pi c}+\frac{c\pi T^{2}}{12H\Lambda _{0}}
\end{eqnarray}%
with $\Lambda _{0}=c\sqrt{\frac{p}{Hc}-\frac{1}{4}}$. After a lengthy
algebra, we find a universal leading order of temperature corrections to the
free energy
\begin{equation}
F=E_{0}-\frac{\pi CT^{2}}{6}\left( \frac{1}{v_{b}}+\frac{1}{v_{f}}\right) ,
\label{LL}
\end{equation}%
where the two velocities in strong repulsive regime are given by
\begin{eqnarray}
v_{s} &=&\frac{4\pi ^{2}n}{3\gamma }\sin (\pi \alpha ),  \notag \\
v_{f} &=&2\pi n\left( 1-\frac{4}{\gamma }\left( \pi \alpha +\sin (\pi \alpha
)\right) \right) .
\end{eqnarray}%
The parameter $\alpha $ is determined by the relation $\alpha \approx
n_{b}/n $ for a small $H\ll 1$. The ground state energy is given by
\begin{equation}
E=\frac{1}{3}n^{3}\pi ^{2}\left( 1-\frac{4}{\gamma }\left( \frac{1}{2}+\frac{%
\sin (\pi \alpha )}{\pi }+(\frac{1}{2}-\alpha )\cos (\pi \alpha )\right)
\right) .
\end{equation}

The TLL is maintained under a crossover temperature at which the linear
temperature-dependent entropy breaks down, see Fig. \ref{fig:S}. The exact
analytic expression of thermodynamic functions (\ref{LL}) from the usual TLL
description is only accurate for a limited range of temperatures and
density. However, the TLL description is incapable of describing quantum
criticality for it does not contain the right fluctuations in the critical
regime. The equation of state (\ref{P}) contains proper universal scaling
functions which control full thermodynamical properties in the quantum
critical regimes. Near critical point, the thermal dynamical properties can
be cased into universal scaling forms, for example, (\ref{us1}) and (\ref%
{us2}). There exists no longer free fermions in the $TLL_{F}$ phase in Fig.%
\ref{fig:S} due to the presence of exponentially small number of bosons at
finite temperatures. They behave like a TTL for temperatures below the
crossover temperatures (white triangles and squares). The $TLL_{M}$ stands
for a two-component TLL of the mixture of bosons and fermions described by (%
\ref{LL}).

\begin{figure}[tbp]
\includegraphics[width=1.0\linewidth]{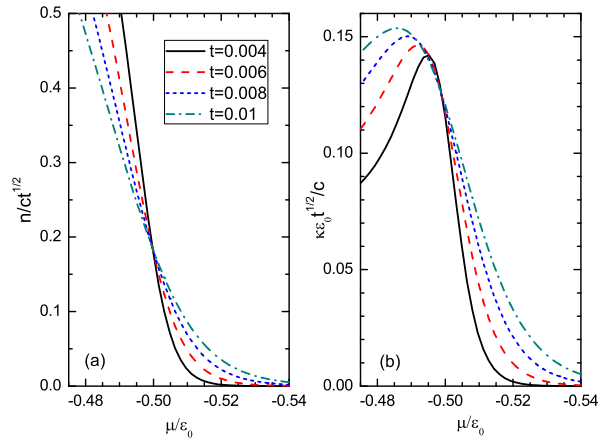}
\caption{(Color online) The density $n_{b}$ and compressibility $\protect%
\kappa _{b}$ vs chemical potential $\protect\mu $ for $H/\protect\varepsilon %
_{0}=-1$ at different temperatures. The curves intersect at the critical
point $\protect\mu =H/2$, i.e. the phase boundary between the pure boson
phase and the vacuum. Here $t=T/\protect\varepsilon _{0}$.}
\label{f2}
\end{figure}

It is straightforward to work out the critical exponents for the
phase transition from vacuum into the the free fermions and into
the Bose gas phase i.e. $d=1$, $z=2$ and $\nu =1/2$, see
\cite{Zhou,Guan,GuanJPA}. Using the TBA equations
(\ref{TBA_nonlinear}), we verify the scaling behavior of the
mixture for the phase transitions from vacuum into the pure Bose
state
in figure \ref{f2} and from vacuum into the pure Fermi state in figure \ref%
{f3}. We see that both the density and compressibility at different
temperatures intersect at the critical point. For both cases, the
compressibility always evolves a round peak as temperature decreases. This
indicates the density of state changes when the phase transition occurs. As
the temperature approaches the limit $T\rightarrow 0$, the compressibility
tends to divergence. 
\begin{figure}[tbp]
\includegraphics[width=1.0\linewidth]{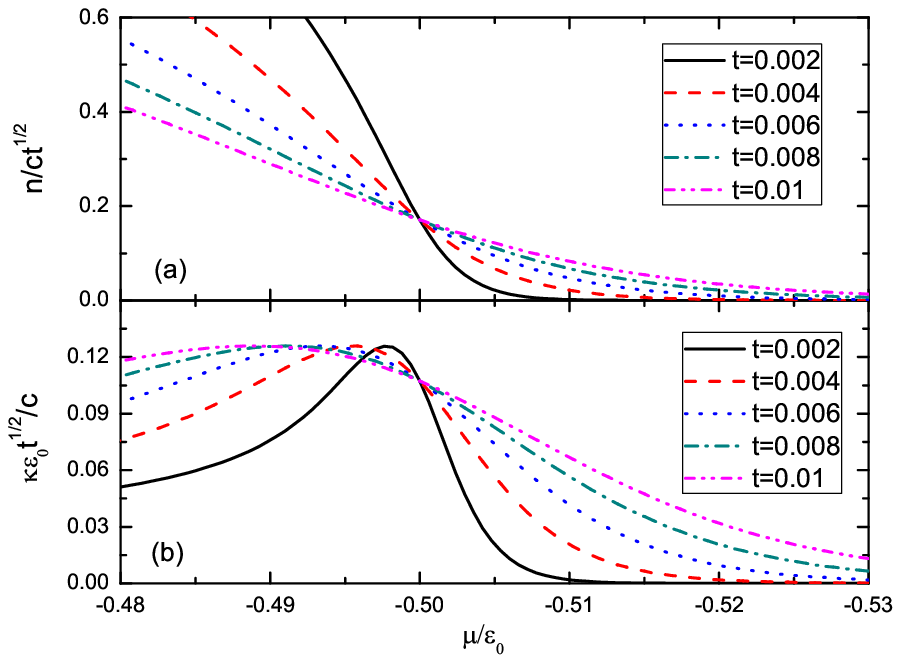}
\caption{(Color online) The density $n_{f}$ and compressibility $\protect%
\kappa _{f}$ vs chemical potential $\protect\mu $ for $H/\protect\varepsilon %
_{0}=1$ at different temperatures. The curves intersect at the critical
point $\protect\mu =-H/2$, i.e. the phase boundary between the pure fermion
phase and the vacuum. Here $t=T/\protect\varepsilon _{0}$.}
\label{f3}
\end{figure}

\begin{figure}[tbp]
\includegraphics[width=1.0\linewidth]{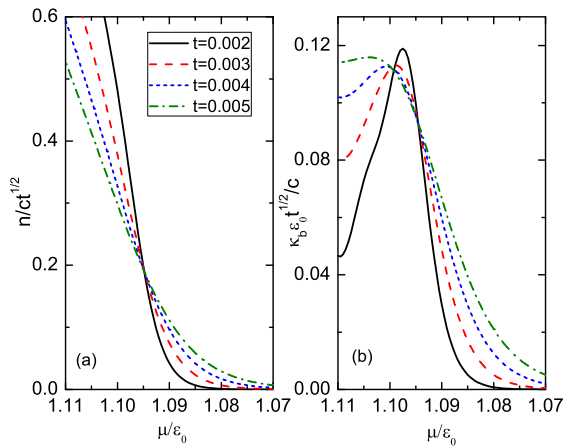}
\caption{(Color online) The density $n_{b}$ and compressibility $\protect%
\kappa_{b}$ vs chemical potential $\protect\mu $ for $H/\protect\varepsilon %
_{0}=1$ at different temperatures. In this setting, the intersection nature
can map out the zero temperature phase boundary for the phase transition
from the free fermions into the mixture of bosons and fermions. Here $t=T/%
\protect\varepsilon _{0}$. }
\label{f4}
\end{figure}

Comparing the density, which is numerically obtained from the pressure (\ref%
{energy}), with the universal scaling form of Eq.(\ref{us1}), we can extract
the critical exponent $z=2$ and the correlation length exponent $\nu =1/2$
with the dimensionality $d=1$. Here we numerically demonstrate the universal
scaling behavior of the mixture of bosons and fermions in Figs \ref{f2}, %
\ref{f3} and \ref{f4}. For practical convenience, here we have chosen the
density of bosons $n_{b}$ to demonstrate the intersections for the phase
transitions from vacuum into the phase of pure bosons and from the phase of
fermions into the mixture, where $n_{b}$ does not have a background near the
transition points, see Figs. \ref{f2}, and \ref{f4}. For a fixed effective
magnetic field $H$, we see that by a proper temperature scaling the density
curves $n_{b}(T)/\sqrt{T}$ at different temperatures intersect at the points
$\mu _{c}$. In the Fig. \ref{f4}, we display the scaled density
distributions for different temperatures by numerically solving the TBA
equations (\ref{TBA_nonlinear})-(\ref{particle_density}). It is also seen
that the compressibility curves at different temperatures intersect at the
critical point $\mu _{c}$. The compressibility tends to divergent as the
temperature tends to zero. It evolves to a round peak at low temperature due
to the change of the density of states around the critical points. The
quantum criticality near the critical points associating the phase
transitions from vacuum into the phase of pure bosons and the phase of pure
fermions reveals a subtle difference in thermodynamical properties, see Figs %
\ref{f2} and \ref{f4}.

\begin{figure}[tbp]
\includegraphics[width=1.0\linewidth]{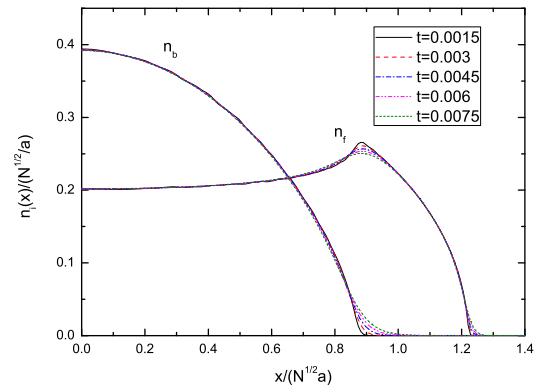}
\caption{(Color online) The density distribution at different temperatures
with $N^{t}a_{1D}^{2}/a^{2}=1$ and $\protect\alpha =0.5$ .}
\label{f5}
\end{figure}

\begin{figure}[tbp]
\includegraphics[width=1.0\linewidth]{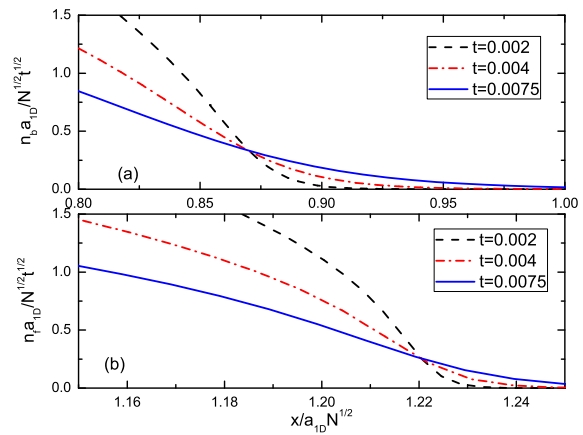}
\caption{(Color online) The densities $n_{b}$ and $n_{f}$\ vs normalized
position for $Na_{1D}^{2}/a^{2}=1$ and $\protect\alpha =0.5$ at different
temperatures. The density curves intersect at the critical point that maps
out the phase boundaries of the trapped gas. Here $t=T/\protect\varepsilon %
_{0}$.}
\label{f6}
\end{figure}

\section{Quantum criticality in the harmonic trap}

\label{Section4} In experiment with cold atoms, the 1D quantum gas is
realized by tightly confining the atomic cloud in two (radial) dimensions
and weakly confining it along the axial direction in an external harmonic
trap. For the mixture of bosons and fermions in a harmonic trap, we can
calculate its density distribution profiles by evaluating the
thermodynamical dynamics within the local density approximation. According
to the local density approximation, the system reaches local equilibrium in
each small intervals around each point $x$ in the external trap. The density
distribution of the trapped gas is then obtained via the local equation of
state \cite{Imambekov,Dunjko}. Within the local density approximation, the
chemical potentials in the equation of state (\ref{energy}) as well as in
the TBA equations are replaced by the local chemical potentials given by
\begin{eqnarray}
\mu _{b}\left( x\right) &=&\mu _{b}\left( 0\right) -V_{b}\left( x\right) ,
\label{LDA1} \\
\mu _{f}\left( x\right) &=&\mu _{f}\left( 0\right) -V_{f}\left( x\right) .
\label{LDA2}
\end{eqnarray}%
Here the external potential is defined as $V_{b}\left( x\right) =V_{f}\left(
x\right) =m\omega ^{2}x^{2}/2$ with harmonic frequency $\omega $ and the
characteristic length for the harmonic trap is $a=\sqrt{\hbar /m\omega }$.
For this case, equations (\ref{LDA1}) and (\ref{LDA2}) can be alternatively
represented as
\begin{equation*}
\mu \left( y\right) /\varepsilon _{0}=\mu \left( 0\right) /\varepsilon
_{0}-y^{2}
\end{equation*}%
for a fixed $H$, where the dimensionless coordinate is given by $%
y=x/(a^{2}c) $. From the Bethe ansatz equations, the dimensionless density $%
n_{b}/c$ and $n_{f}/c$ can be obtain for fixed dimensionless chemical
potential $\mu _{b}/\varepsilon _{0}$ and $\mu _{f}/\varepsilon _{0}$.\ The
total particle number $N$ is obtained from%
\begin{equation*}
\frac{Na_{1D}^{2}}{a^{2}}=4\int_{-\infty }^{\infty }\frac{n_{b}\left(
y\right) +n_{f}\left( y\right) }{c}dy
\end{equation*}%
with the 1D scattering length $a_{1D}=-2/c$. We define a polarization rate
between the Bose particle number and the total particle number $\alpha
\equiv N_{b}/N$. For fixed $\mu \left( 0\right) /\varepsilon _{0}$ and $%
H/\varepsilon _{0}$, we can calculate $Na_{1D}^{2}/a^{2}$ and
$\alpha $. In the presence of the confined potential, the length
scale of the system at quantum criticality is still much smaller
than the trap size. Therefore, the critical behavior of the
homogeneous gas can be mapped out by the density profiles of gas
at finite temperatures \cite{Zhou}.

We fix $Na_{1D}^{2}/a^{2}$ and $\alpha $ in the trapped gas of the mixture,
the density profiles reveal a universal scaling behavior of the gas. Fig. {%
\ref{f5}} shows the density profiles of bosons and fermions in the
harmonic trap for $Na_{1D}^{2}/a^{2}=1$ and $\alpha =N_{b}/N=0.5$
at different temperatures. Here we find that bosons and fermions
coexist in the trap center companied by the phase of pure fermions
at the edges. We further demonstrate how to map out the
zero-temperature phase boundaries from the
density profiles of the trapped gas at finite temperatures. In Fig. {\ref{f6}%
}(a) and (b), we demonstrate the scaled density distributions of bosons and
fermions. It is clearly seen that the scaled density curves for different
temperatures intersect at the critical point of the trapped gas. Thus the
critical point, separating the mixture of bosons and fermions from the phase
of pure fermions, is mapped out. Similarly, the density curves of fermions
intersect at the critical point that maps out the phase boundary for the
phase transition from vacuum into the phase of free fermions.

\section{Conclusion}

\label{Section5} In summary, we have studied the phase diagram,
universal TLL and quantum criticality of the 1D Bose-Fermi mixture
by means of the TBA equations. We have derived the equation of
state and universal TLL thermodynamics of the model for strong
repulsion. We have proved that the low energy physics of the
Bose-Fermi mixture are described by a two-component TTL. Universal
scaling behavior of thermodynamical properties at quantum
criticality provides a physical origin of quantum critical
phenomena. Furthermore, the quantum criticality of the Bose-Fermi
mixture in an harmonic trap has been studied within the local
density approximation. It turns out that the phase diagram and
critical properties of the bulk system can be mapped out from the
density profiles of the trapped mixture gas at finite
temperatures. Our exact results can help with experimental study
of quantum critical phenomena in a 1D harmonic trap.

\textit{Acknowledgments.---} This work was supported by the
National Program for No. 2011CB921700 Basic Research of MOST, NSF
of China under Grants No. 10821403, No. 11174360 and No. 10974234,
and 973 grant. The work of X.-W.G has been partially supported by
the Australian Research Council. YZ is also supported by 973
Program under Grant No. 2011CB921601, and the Program for NCET.

\end{document}